\documentclass[fleqn,twoside]{article}

\newcommand{\heads}[2]{\markboth{\protect\small\it #1}{\protect\small\it #2}}
\newcommand{\Acknow}[1]{\subsection*{\normalsize Acknowledgement} #1}
\def\noi{\noindent}
\newcommand{\foom}[1]{\protect\footnotemark[#1]}
\newcommand{\email}[2]{\footnotetext[#1]{e-mail: #2}
        \addtocounter{footnote}{1}}
\newcommand{\Title}[1]{\noi {\uppercase{\Large #1}} \\}
\newcommand{\Author}[2]{\noi{\large\bf #1}\\[2ex]\noindent{\it #2}\\}
\newcommand{\Abstract}[1]{\vskip 2mm \begin{center}
        \parbox{16.4cm}{\small\noi #1} \end{center}\medskip}

\newcommand{\Arthead}[5]{ \setcounter{page}{#4}\thispagestyle{empty}\noi
    \unitlength=1pt \begin{picture}(500,40)
        \put(0,58){\shortstack[l]{\small\it Gravitation \& Cosmology,
                        \small\rm Vol. #1 (#2), No. #3, pp. #4--#5\\
\small
Proceedings of the International Conference on Gravitation, Cosmology,
Astrophysics and Nonstationary Gas Dynamics,\\
\small
Dedicated to Prof. K.P. Staniukovich's 90th birthday, Moscow, 2-6 March 2006
} }
    \end{picture}
     }          

\heads{Yu. V. Pavlov}
{Duffin-Kemmer-Petiau equation with nonminimal coupling to curvature}

\begin{document}

\twocolumn[
\Arthead{12}{2006}{2-3 (46-47)}{205}{208}
\setcounter{page}{1}

\Title{Duffin-Kemmer-Petiau equation \\[5pt]
with nonminimal coupling to curvature}

\Author{Yu.V. Pavlov\foom 1}
{Institute of Mechanical Engineering, Russian Acad. Sci.,
61 Bolshoy Pr. V.O., St.\,Petersburg, 199178, Russia;
\newline
A.\,Friedmann Laboratory for Theoretical Physics, St.\,Petersburg, Russia}

\Abstract
{The generalized Duffin-Kemmer-Petiau equation in curved
space-time is proposed for non-minimal coupling to the curvature
and external fields.
The corresponding scalar and vector fields equation are found.
Equations are presented, which are equivalent to those of a scalar field
with conformal coupling and electromagnetic field with non-minimal
coupling to the curvature.
The gauge-invariant Duffin-Kemmer-Petiau equation with non-minimal
coupling is given.}

] 
\email 1 {yuri.pavlov@mail.ru}

\section{\large Introduction}

The Duffin-Kemmer-Petiau (DKP) equation is a first-order relativistic
wave equation for spin 0 and 1 bosons \cite{PDK}.
Recently there has been an increasing interest to DKP theory in external
fields and curved space-time (see~\cite{LPT} and references there).
For minimal coupling to the curvature or an external vector field,
the DKP equation in the scalar case is equivalent to
the corresponding Klein-Gordon-Fock equation~\cite{LPT,FP00}.
For the vector case, the DKP equation with minimal coupling is equivalent
to the Maxwell or Proca equations.

In the DKP formalism, a wave function is multicomponent.
That is why the simplest non-minimal interactions with external
fields have a more complicated structure than in usual formalism.
It has applications in describing of interactions of mesons with
nuclei~\cite{CK}, for studies of pionic atoms~\cite{K86} etc.
The question on non-minimal coupling to the curvature in the DKP formalism
was not considered earlier, with the exception of the work~\cite{CQG}.
However, in this work, the DKP equation with conformal coupling was written
only using an auxiliary field obeying contradictory conditions:
this field is vector for general coordinate transformation and
is constructed only from the metric tensor and its derivatives.

In the present work, a generalized DKP equation is introduced, with
non-minimal coupling to the curvature and external fields,
and the corresponding scalar and vector field equations are found.
We use the system of units where $\hbar =c=1$.
The signs of the curvature tensor and the Ricci tensor are
chosen such that
$ R^{\, i}_{\ jkl} = \partial_{\,l}  \Gamma^{\, i}_{\, jk} -
\partial_k  \Gamma^{\, i}_{\, jl} +
\Gamma^{\, i}_{\, nl} \Gamma^{\, n}_{\, jk} -
\Gamma^{\, i}_{\, nk} \Gamma^{\, n}_{\, jl}\ $
and  $\, R_{ik} = R^{\, l}_{\ ilk}$\,, $R=R^k_k$,
where $\Gamma^{\, i}_{\, jk}$ are Christoffel symbols.

\section{\large DKP formalism}

The DKP equation for a field with mass $m$ is given by
    \begin{equation}
(i \beta^a \partial_a - m) \psi = 0\,,
\label{eq1}
\end{equation}
where the matrices $\beta^a$ obey the algebraic relations
    \begin{equation}
\beta^a \beta^b \beta^c + \beta^c \beta^b \beta^a =
\beta^a \eta^{bc} + \beta^c \eta^{ba},
\label{eq2}
\end{equation}
and $\eta^{ab}$ is a constant diagonal matrix (in particular,
${\rm diag}(1,-1,\ldots,-1)$).
For the space-time dimension $N=4$, the DKP algebra~(\ref{eq2}) has
5- (corresponding to spin 0), 10- (corresponding to spin 1) and
1-dimensional (trivial) irreducible representations.

Each component of the free massive DKP field obeys the Klein-Gordon-Fock
equation, because
    \begin{equation}
d(\partial) (i \beta^a \partial_a - m) =
- (\partial_a \partial^a +m^2) I \,,
\label{dLm}
\end{equation}
where $I$ is the identity matrix and
    \begin{equation}
d(\partial) = m I + i \beta^a \partial_a + \bigl(
2 \eta^{ab} \!- \beta^a \beta^b \!- \beta^b \beta^a \bigr)
\frac{\partial_a \partial_b}{2m}\,.
\label{dpm}
\end{equation}
The DKP equation for the massless case was written by
Harish-Chandra~\cite{HCh}
    \begin{equation}
i \beta^a \partial_a \psi - \gamma \psi = 0\,.
\label{eqm0}
\end{equation}
Here $\gamma$ is a singular matrix satisfying
    \begin{equation}
\beta^a \gamma + \gamma \beta^a = \beta^a, \ \ \
\gamma^2 = \gamma \,.
\label{eqgamma}
\end{equation}
If $\gamma$ is a solution to~(\ref{eqgamma}), then $1-\gamma$ is also one.
The following theorem takes place~\cite{HCh}: for any irreducible set of
matrices $\beta^a$ there exist either none or just two such
matrices $\gamma$.

We introduce Umezawa projectors~\cite{Umezawa}, generalized to
$N$-dimensional case, and an arbitrary diagonal matrix $(\eta^{ab})$:
    \begin{equation}
P= {\rm det}(\eta_{ab}) (\beta^0)^2 (\beta^1)^2 \cdots (\beta^{N-1})^2,
\ \  P^a=P \beta^a,
\label{defPPa}
\end{equation}
    \begin{equation}
Q^a = {\rm det}(\eta_{ab})\, (\beta^1)^2 \cdots (\beta^{N\!-1})^2
[ \eta^{a0} - \beta^a \beta^0 ]\,,
\label{RaRab}
\end{equation}
    \begin{equation}
Q^{ab} = Q^a \beta^b \,.
\label{Qab}
\end{equation}

Under the infinitesimal Lorentz transformations
    \begin{equation}
x^{\prime\,a} = \Lambda^a_{\ b} x^b , \ \ \
\Lambda^a_{\ b} = \delta^a_{\ b} + \omega^a_{\ b}\,, \ \ \
\omega_{ab} = - \omega_{ba}
\label{Lmal}
\end{equation}
we have \ \ $ \psi'(x') = S(\Lambda) \psi(x)$,
    \begin{equation}
S(\Lambda) = 1 +\frac{1}{2} \, \omega_{ab} S^{ab}, \ \ \
S^{ab} = \beta^a \beta^b - \beta^b \beta^a ,
\label{Ul}
\end{equation}
and $P\psi$ transforms as scalar, $P^a \psi$ and $Q^a \psi$  as vectors,
and $Q^{ab} \psi$  as antisymmetric tensor of second rank.

The DKP equation for curved space is generalized~\cite{LPT}
analogously to the Fock-Ivanenko-Weyl method~\cite{WFI} for
the Dirac equation.
On a space-time manifold, a sets of N vector fields
$e^{\, i}_{(a)}(x)$ and $e^{(b)\,i}(x)$
(tetrads in the four-dimensional case)
are introduced with the relations
    \begin{equation}
e^{\, i}_{(a)} e_{(b) i} = \eta_{ab} \ , \ \ \ \
e_{(a)}^{\, i} e^{(b)}_{\,i} = \delta_a^{\,b}.
\label{edef}
\end{equation}
The covariant derivative of the DKP field $\psi$ is defined by
    \begin{equation}
\nabla_{\!k} \psi = \bigl( \partial_k + \frac{1}{2} \omega_{k ab} S^{ab}
\bigr) \psi \,,
\label{defDPSI}
\end{equation}
where the ``spin'' connection $\omega_{kab}$ obeys the relations
    \begin{equation}
\omega_{k ab} = e_{(a) l}\, e_{(b)}^{\,j} \Gamma^{\,l}_{\,jk} \!-
e_{(b)}^{\,j} \partial_k e_{(a) j} \,, \ \ \
\omega_{k ab} = - \omega_{k ba}.
\label{omega}
\end{equation}
The DKP equations in curved space-time, for the massive and massless
fields, are written as
    \begin{equation}
i \beta^k \nabla_{\!k} \psi - m \psi = 0\,, \ \ \
i \beta^k \nabla_{\!k} \psi - \gamma \psi = 0,
\label{DKPcurved}
\end{equation}
where $\beta^k=e^{\, k}_{(a)} \beta^a$.
There equations are equivalent to those
of minimally coupled scalar or vector fields~\cite{LPT,CPLT}.

\section{\large Nonminimal coupling to curvature}

Usually, in quantum theory in curved space-time, a scalar field
is considered with the Lagrangian
    \begin{equation}
L(x)= \!\sqrt{|g|} \left[ \partial_i\varphi^*\partial^i\varphi -
(m^2 \!+ V_{\!g}) \varphi^* \varphi - U(\varphi^* \varphi) \right]
\label{Lag}
\end{equation}
and the corresponding equation of motion
    \begin{equation}
\left( \nabla^i \nabla_{\! i} + V_{\!g} + m^2 + U'(\varphi^* \varphi)
\right) \varphi(x)=0 \,,
\label{Eqm}
\end{equation}
where $\nabla_{\! i}$ are covariant derivatives in
the metric $g_{ik}, \ g={\rm det} (g_{ik})$, $U(\varphi^* \varphi)$ is
self-interaction, $V_{\!g}=0$ for minimal coupling to curvature.
    In the case $V_{\!g}=\xi_c R$, where $\xi_c=(N-2)/[4(N-1)]$
(conformal coupling), the massless equation is conformally invariant if
$U=\lambda(\varphi^* \varphi)^{N/(N-2)}$,\ $\lambda={\rm const}$.
    For $V_{\!g}$ of arbitrary form, the third and fourth derivatives
can appear in the metric energy-momentum tensor and the Einstein equations.
However, for the Gauss-Bonnet-type coupling~\cite{Pv4}
    \begin{equation}
V_{\!g} = \xi R + \chi (R_{lmpq} R^{\,lmpq} - 4 R_{lm} R^{\,lm} + R^2)
\label{V}
\end{equation}
the energy-momentum tensor does not contain higher than the second-order
derivatives of the metric.

The electromagnetic field with the tensor
$F^{ik}=\partial^i A^k-\partial^k A^i$ and
a minimal coupling to curvature has the equation
    \begin{equation}
\nabla_{\!i} F^{ik} = 0 \,,
\label{emmin}
\end{equation}
which is conformal invariant if $N=4$.
For a vector field with nonminimal coupling to the curvature,
the equation is often chosen~\cite{NovS}-\cite{Teys} in the form
    \begin{eqnarray}
\nabla_{\!i} \Bigl( (1 - \lambda_1 R) F^{ik} -
\lambda_2 (R^i_n F^{nk} - R^k_n F^{ni}) -   \nonumber   \\
- \lambda_3 R^{iklm} F_{lm} \Bigr)
+ (\xi_1 R + m^2) A^k + \xi_2 R_i^k A^i = 0,
\label{eqII}
\end{eqnarray}
where $\lambda_1, \lambda_2, \lambda_3, \xi_1, \xi_2$ are constants.
The corresponding Lagrangian is
    \begin{equation}
L_{v} = - \frac{1}{4}\, \sqrt{|g|}\, F_{ik} F^{ik} + L_{I} + L_{II} \,,
\label{Lem}
\end{equation}
where
    \begin{eqnarray}
L_{I} = \frac{\sqrt{|g|}}{4} \bigl[ \lambda_1 R F_{ik} F^{ik}
+ 2 \lambda_2 R_{ik} F^{in} F^{k}_{\ n}  +   \nonumber   \\
+\, \lambda_3 R_{iklm} F^{ik} F^{lm} \bigr],
\label{LagI}
\end{eqnarray}
    \begin{equation}
L_{II} = \frac{\sqrt{|g|}}{2} \left[ (\xi_1 R + m^2) A_i A^i +
\xi_2 R_{ik} A^i A^k \right].
\label{LagII}
\end{equation}
The additional terms from~(\ref{LagII}) broke the gauge invariance of
the theory.
The additional terms from~(\ref{LagI}) can produce a lot of new effects:
photons creation in expanding Universe~\cite{NovelloOS90}
(for Eq.~(\ref{emmin}), photons do not create due to conformal
invariance);
polarization dependence of the photon trajectory~\cite{PrMoh};
variation of the speed of light in curved space~\cite{Teys}, etc.

\section{\large  DKP equation with non-minimal coupling}

We consider the generalized DKP equations for non-minimal
coupling to the curvature and external fields:
\begin{eqnarray}
i \beta^k \left( \nabla_{\!k} + i B_k \right)
\Bigl( 1 + \gamma \sum \zeta_{k_1 \cdots\, k_p} \beta^{k_1} \!\! \ldots
\beta^{k_p} \Bigr) \psi -   \nonumber  \\
-\, \alpha \gamma \psi -
\sum V_{k_1 \cdots\, k_q} \beta^{k_1} \!\! \ldots \beta^{k_q} (1-\gamma)
\psi = 0,
\label{DKPnm}
\end{eqnarray}
where $\alpha = {\rm const} \ne 0 $,
$B_k(x)$ is an external vector field,
$ \zeta_{k_1 \cdots\, k_p}(x),\ V_{k_1 \cdots\, k_q}(x) $
($p,q= 0,1,2, \ldots $)
are arbitrary external fields, e.g., $R, \ R_{ik}, \ R_{iklm} $.
To find the scalar equation corresponding to Eq.~(\ref{DKPnm}),
we use the DKP algebra and the following relations:
$P^i \beta^j = g^{ij} P $,
    \begin{equation}
P \nabla_{\!i} \psi = \nabla_{\!i} (P \psi) = \partial_i (P \psi)
= P \partial_i \psi \,,
\label{Pdpsi}
\end{equation}
    \begin{equation}
P^i \nabla_k \psi = \nabla_k (P^i \psi) =
\partial_k (P^i \psi) + \Gamma^{\,i}_{\,lk} P^l \psi \,,
\label{Pidkp}
\end{equation}
    \begin{eqnarray}
P \sum V_{k_1 \cdots\, k_q} \beta^{k_1} \! \ldots \beta^{k_q} =
\nonumber   \\
= \sum \limits_{\rm even}
V_{k_2 \ \cdots \ k_q}^{\ k_2}{\!\!\mathstrut}^{k_q} P +
\sum \limits_{\rm odd}
V_{k_1 \ \cdots \ k_{q-2}}^{\ k_1 \ \ \ \ k_{q-2}}{\mathstrut
\!\!}_{k_q} P^{k_q},
\label{PSumV}
\end{eqnarray}
    \begin{eqnarray}
P^k \sum \zeta_{k_1 \cdots\, k_p} \beta^{k_1} \! \ldots \beta^{k_p} =
\nonumber  \\
= \sum \limits_{\rm even} \zeta^{k \ \, k_2 \ \ \ \ \
k_{p-2}}_{\ k_2 \ \cdots \ k_{p-2} \ \ k_p} P^{k_p} +
\sum \limits_{\rm odd} \zeta^{k \ \, k_3}_{\ k_3 \
\cdots \ k_p}{}^{\! k_p}  P,
\label{Pzeta}
\end{eqnarray}
where $ P^i = e^{\,i}_{(a)} P^a $,
$ \sum \limits_{\rm even} $ and $ \sum \limits_{\rm odd} $
denote summation with even and odd number of indices.
Multiplying~(\ref{DKPnm}) by $(1-\gamma)P$ and $(1-\gamma)P^l$
and taking into account~(\ref{Pdpsi})--(\ref{Pzeta}),
and the equalities following from (\ref{eqgamma})
    \begin{equation}
\gamma (1-\gamma) = 0, \ \ \
\gamma \beta^k = \beta^k (1-\gamma), \ \ \
\gamma \beta^i \beta^k = \beta^i \beta^k \gamma ,
\label{gammaP}
\end{equation}
    \begin{equation}
\gamma P = P \gamma \,, \ \ \ \
\gamma P^k = P^k (1-\gamma),
\label{gamP}
\end{equation}
we obtain
    \begin{eqnarray}
(1-\gamma) \biggl[ i D_k \Bigl( \Bigl( \delta^k_l +
\sum \limits_{\rm even} \zeta^{k \ \, k_2}_{\ k_2 \ \cdots \ l} \Bigr)
P^l +     \nonumber  \\
+ \sum \limits_{\rm odd} \zeta^{k \ \, k_3 \ \ k_p}_{\ k_3 \, \cdots \, k_p}
P \Bigr) -
\sum \limits_{\rm even} V_{k_2 \ \cdots \, k_q}^{\ k_2 \ \ \ k_q} P
\biggr] \psi = 0 \,,
\label{DKPgP}
\end{eqnarray}
    \begin{equation}
(1-\gamma) P^l \psi = \frac{1-\gamma}{\alpha} \, \Bigl[ i D^l -
\sum \limits_{\rm odd} V^{l \ k_3 \ \ \ k_q}_{\ k_3 \ \cdots \, k_q}
\Bigr] P \psi \,,
\label{DKPgPl}
\end{equation}
where \ $D_k = \nabla_{\!k} + i B_k $.
Substituting~(\ref{DKPgPl}) in (\ref{DKPgP}), one obtains
an equation for the scalar component $\phi \equiv (1-\gamma) P \psi$:
    \begin{eqnarray}
\hspace{-7pt}
\biggl\{\! D_k \biggl[ \Bigl( g^{kl} \!+\! \sum \limits_{\rm even}
\zeta^{k \ k_3 \ \ \ k_{p-\!1}\, l}_{\ k_3 \ \cdots \, k_{p-\!1}} \Bigr)
\Bigl(\! D_l \!+ i\! \sum \limits_{\rm odd}
V_{l\, k_3 \ \cdots \, k_q}^{\ \ k_3 \ \ \ k_q} \Bigr) \biggr] \!-
\hspace{-17pt}
\nonumber \\
- i \alpha D_k \Bigl( \sum \limits_{\rm odd}
\zeta^{k \ k_3 \ \ \ k_p}_{\ k_3 \ \cdots \, k_p} \Bigr) +
\alpha \sum \limits_{\rm even} V_{k_2 \ \cdots \, k_q}^{\ k_2 \ \ \ k_q}
\biggr\} \phi = 0 \,.
\label{DDP}
\end{eqnarray}
For a scalar representation of the DKP algebra, this equation is
equivalent to~(\ref{DKPnm}).
The DKP equation corresponding to Eq.~(\ref{Eqm}) may be written as
    \begin{equation}
\hspace{-5pt} i \beta^k \nabla_{\!k} \psi - \alpha \gamma \psi -\!
\frac{1}{\alpha} \bigl( m^2 \!+ V_{\!g} P +
U' P \bigr) (1 \!- \gamma) \psi = 0,\
\label{scex}
\end{equation}
where $ U' = U'(\alpha^{-1} \psi^{+} P^+ (1-\gamma^+) (1-\gamma) P \psi ) $.
If $m \ne 0 $, then the choice $\alpha = m$ gives identical dimensions of
different components of the DKP field.
For the massless case we take $\alpha=1$, then the dimensions of
$P \psi$ and $P^k \psi$ are different.

For spin 1 we take into account that
    \begin{equation}
Q^{ik} = - Q^{ki} \ , \ \ \ \ \
Q^i \beta^k \beta^l = Q^{ik} \beta^l = Q^i g^{kl} - Q^k g^{il}
\label{Qikl}
\end{equation}
    and introduce the tensors
$ E^{lk}_{\ \, mn} $, \ $ O^{lk}_{\ \, n} $, \
$ H^{l}_{\ n} $, \ $ G^{lk}_{\ \, n} $:
    \begin{eqnarray}
Q^{lk} \Bigl( 1 + \sum \limits_{\rm even} \zeta_{k_1 \cdots \, k_p}
\beta^{k_1} \cdots \beta^{k_p} \Bigr) = E^{lk}_{\ \, mn} Q^{mn},
\\
Q^{lk} \sum \limits_{\rm odd} \zeta_{k_1 \cdots \, k_p}
\beta^{k_1} \cdots \beta^{k_p} = O^{lk}_{\ \, n} Q^{n},
\\
Q^{l} \sum \limits_{\rm even} V_{k_1 \cdots \, k_q}
\beta^{k_1} \cdots \beta^{k_q} = H^{l}_{\ n} Q^{n},
\\
Q^{lk} \sum \limits_{\rm odd} V_{k_1 \cdots \, k_q}
\beta^{k_1} \cdots \beta^{k_q} = G^{lk}_{\ \, n} Q^{n} .
\label{QZQV}
\end{eqnarray}
Multiplying (\ref{DKPnm}) by $(1-\gamma) Q^l$ and $(1-\gamma) Q^{ln}$,
one obtains
    \begin{equation}
(1\!-\gamma) \biggl[ i D_k \Bigl(\! E^{lk}_{\ \ mn} Q^{mn} +
O^{lk}_{\ \ n} Q^n\! \Bigr) - H^l_{\ n} Q^n \biggr] \psi = 0,
\label{gQl}
\end{equation}
    \begin{eqnarray}
(1-\gamma) Q^{ln} \psi =
\nonumber  \\
= \frac{1-\gamma}{\alpha} \, \Bigl[ i
\left( D^n Q^l - D^l Q^n \right) - G^{ln}_{\ \ m} Q^m \Bigr] \psi.
\label{gQln}
\end{eqnarray}
    We denote $A^k = (1-\gamma) Q^k \psi $,
$$   
F^{ik} = \left( D^i Q^k \!- D^k Q^i \right) (1 \!- \gamma) \psi
= D^i A^k \!- D^k A^i,
$$   
    and substitute (\ref{gQln}) in (\ref{gQl}).
As a result, we have
    \begin{eqnarray}
D_k \left( E^{lk}_{\ \ mn} F^{nm} \right) +
\nonumber   \\
+ \left[ i D_k \left( E^{lk}_{\ \ mn} G^{mn}_{\ \ p} -
\alpha O^{lk}_{\ \ p} \right) + \alpha H^l_{\ p} \right] A^p = 0.
\label{DF}
\end{eqnarray}
For a vector representation of the DKP algebra, this equation is
equivalent to~(\ref{DKPnm}).
In particular, for a vector representation, the following DKP equation:
    \begin{eqnarray}
i \beta^k \nabla_{\!k} \biggl( 1 - \gamma \Bigl( \lambda_1 R + \lambda_2
R_{mn} \beta^m \beta^n +  \hspace{14pt}
\label{DDPemf}   \\ +
\frac{\lambda_3}{2} \left( R_{mn} \beta^m \beta^n + R_{mnpq} \beta^m \beta^p
\beta^n \beta^q \right) \Bigr) \biggr) \psi - \alpha \gamma \psi -
\hspace{-17pt} \nonumber    \\[1mm] -\, \frac{1}{\alpha} \Bigl( m^2 + \xi_1
R + \xi_2 (R - R_{mn} \beta^m \beta^n) \Bigr) (1-\gamma) \psi =0
\hspace{-17pt}  \nonumber
\end{eqnarray}
reproduces Eq.~(\ref{eqII}) with non-minimal coupling.

The DKP equation
    \begin{equation}
i \beta^k \nabla_{\!k} \Bigl( 1 + \gamma \sum \zeta_{k_1 \cdots\, k_{2n}}
\beta^{k_1} \! \ldots \beta^{k_{2n}} \Bigr) \psi - \gamma \psi = 0
\label{DKPginv}
\end{equation}
is invariant under the transformation
    \begin{equation}
\psi \to \psi + (1-\gamma) \Phi \,,  \ \ \ {\rm if} \ \ \
\gamma \beta^k \nabla_{\!k} \Phi=0\,.
\label{gin}
\end{equation}
In a vector representation, such transformations correspond to
addition of derivatives of some scalar function to the components
$A_k = (1-\gamma) Q_k \psi $ \cite{CPLT}, i.e., Eq.~(\ref{DKPginv})
corresponds to the non-minimally coupled gauge-invariant vector field.

The DKP equation of the form
    \begin{equation}
i \beta^k \nabla_{\!k} \psi - \gamma \psi - \xi_c R P (1 - \gamma) \psi= 0
\label{sconm0}
\end{equation}
describes the conformally invariant scalar field in a scalar
representation of the DKP algebra and the minimally coupled
electromagnetic field~(\ref{emmin}), for a vector representation $(P=0)$.

For DKP algebra representations in which
    \begin{equation}
(\beta^0)^+ =\beta^0 \ , \ \ \ (\beta^\nu)^+ = - \beta^\nu
\ , \ \ \ \ \gamma^+ = \gamma \,,
\label{sopr}
\end{equation}
where $\nu=1, \ldots , N-1$, the Lagrangian for the DKP equation
can be written similarly to that for the Dirac equation.
For examples, the Lagrangian corresponding to Eq.~(\ref{scex}) is
    \begin{eqnarray}
L = \sqrt{|g|}\, \Bigl[ i \overline{\psi} \gamma \beta^k \nabla_{\!k}
\psi - i \nabla_{\!k} \overline{\psi} \beta^k \gamma \psi -
\alpha \overline{\psi} \gamma \psi - \hspace{4pt}
\label{Lmls}   \\
-\, \alpha^{\!-1} \overline{\psi}(m^2 \!+ V_{\!g} P) (1\!- \gamma) \psi -
U \bigl(  \alpha^{\!-1} \overline{\psi} (1\!- \gamma) P \psi  \bigr) \Bigr],
\hspace{-14pt}  \nonumber
\end{eqnarray}
where \ $ \overline{\psi} = \psi^+ \eta^0 $ is the DKP conjugate function,
\ $ \eta^0 = 2(\beta^0)^2/\eta^{00} - 1 $,
    \begin{equation}
\nabla_{\! k} \overline{\psi} = \partial_k \overline{\psi} - \frac{1}{2}
\omega_{kab} \overline{\psi} S^{ab}.
\label{npsi}
\end{equation}

Thus the DKP formalism is an equivalent form for the description of scalar
and vector fields with various types of coupling to the curvature
and external fields
(see Eqs.~(\ref{DKPnm}), (\ref{DDP}), (\ref{DF}) and the particular
cases~(\ref{scex}), (\ref{DDPemf}), (\ref{DKPginv}), (\ref{sconm0})).
Taking into account a possible non-minimal coupling of the scalar and
vector fields to curvature may be important for the early Universe.
The questions concerning the type of coupling to the curvature pertain to
experiment.
The generalized DKP equations can be used in those problems,
where the general covariance is required and the formalism
of first-order differential equations is preferable.

\Acknow {The author is grateful to Prof. A.A.\,Grib for helpful discussion.
    This work was supported by the Ministry of Science and Education of
Russia, grant RNP.2.1.1.6826.}

\end{document}